# Mechanical control of physical properties in the van der Waals ferromagnet $Cr_2Ge_2Te_6$ via application of electric current

Hengdi Zhao[1], Yifei Ni[1], Bing Hu[1,2], Sabastian Selter[3], Saicharan Aswartham[3], Yu Zhang[1], Bernd Büchner[3,4], Pedro Schlottmann[5] and Gang Cao[1*]

[1]Department of Physics, University of Colorado at Boulder, Boulder, CO 80309, USA

[2]School of Mathematics and Physics, North China Electric Power University, Beijing 102206, China

[3]Institute for Solid State Research, Leibniz IFW Dresden, Helmholtzstrasse 20, 01069 Dresden, Germany

[4]Institute of Solid State and Materials Physics, Technische Universität Dresden, 01062 Dresden, Germany

[5]Department of Physics, Florida State University, Tallahassee, FL 32306, USA

$Cr_2Ge_2Te_6$ is a van der Waals ferromagnet with a Curie temperature at 66 K. Here we report a swift change in the magnetic ground state upon application of small DC electric current, a giant yet anisotropic magnetoelectric effect, and a sharp, lattice-driven quantum switching manifested in the I-V characteristic of the bulk single-crystal $Cr_2Ge_2Te_6$. At the heart of these observed phenomena is a newly uncovered, strongly anisotropic magnetoelastic coupling that enables strongly anisotropic responses of the lattice to application of electric current and/or magnetic field, thus the exotic phenomena in $Cr_2Ge_2Te_6$. Such a rare mechanical tunability in the magnetic semiconductors promises tantalizing prospects for unique functional materials and devices.

*gang.cao@colorado.edu

Chalcogenides as electronic materials have been studied for decades. From large thermoelectric effect, superconductivity, photovoltaic effect, to topological behavior, and most recently, novel colossal magnetoresistance [e.g.,1-4], this class of materials is constantly surprising us with exotic phenomena. In recent years, the realization of novel physics inherent in two-dimensional (2D) semiconductors first exemplified by graphene in 2004 [5] has provided impetus for a burgeoning group of studies of a large array of 2D transition-metal chalcogenides $A_2B_2X_6$ (A = transition metal, B = Si, Ge or P and X = S, Se or Te) [e.g., 6-12]. These chalcogenides commonly feature a narrow band gap (< 2 eV) and significant spin-orbit interactions. Among them is the title compound $Cr_2Ge_2Te_6$, which was first reported in 1995 [13]. It is a van der Waals ferromagnet with a Curie temperature, $T_C$, at 66 K, which is a result of a competition between the AFM and FM interactions between the $3d$ and $5p$ electrons conducting via the planar Cr-Cr pathway and the 90°-interplanar Cr-Te-Cr pathway (Goodenough-Kanamori rules [14,15]), respectively [13]. The renewed interest in this and other related chalcogenides such as $Cr_2Si_2Te_6$ [6,16] arises in part from the realization of ferromagnetic order in atomic layers of these materials, such as $Cr_2Ge_2Te_6$ [7], in which the magnetic anisotropy overpowers thermal fluctuations, stabilizing the long-range magnetic order [7] (Mermin-Wagner theorem).

The magnetic anisotropy was the main motive of the most recent studies, e.g., through scanning magneto-optic Kerr microscopy (MOKE) [7], scanning tunneling spectroscopy [9], anisotropy in the magnetoresistance [10], magnetization and specific heat measurements [11], and electron spin resonance (ESR) and ferromagnetic resonance (FMR) [10,17].

Unlike transition-metal oxides where oxygen has the strong electronegativity giving oxides a tendency to have the ionic bonding, the chalcogenides, especially the tellurides, host instead a strong competition between metal-ligand and ligand-ligand bonding, which becomes a driver of structural and physical properties in these materials (note that the electronegativity decreases following the order of O, S, Se and Te, the ionic bonding is no longer as significant in heavier chalcogenides) [18]. Such structural characteristics render a high susceptibility of the ground state to external stimuli that couple to the lattice. This tunability promises a rich phenomenology and indicates that the *bulk* tellurides are strongly under-investigated as compared to other material classes.

Here we report a drastic change in the magnetic ground state upon application of a small DC electric current, a giant yet anisotropic magnetoelectric effect and a sharp quantum switching evident in the I-V characteristic in the bulk single-crystal $Cr_2Ge_2Te_6$. In essence, a combination of the single-crystal x-ray diffraction as functions of both temperature and electric current with the measurements of magnetodielectric effect reveals a strong *magnetoelastic* coupling that occurs only along the *a* axis. Application of the electric current expands the *c* axis but shortens the *a* axis. This anisotropic change in the lattice disproportionally enhances the direct antiferromagnetic (AFM) exchange interaction operating via the planar Cr-Cr bond distance, thus destabilizes the precarious balance between the AFM interaction and the ferromagnetic (FM) superexchange interaction via the 90°-interplanar Cr-Te-Cr pathway, eventually destroying the native FM state in a critical regime of current density, $0.31 < J_C < 0.55 A/cm^2$. Above $J_C$, a paramagnetic state emerges with a finite negative Curie-Weiss temperature, indicating a dominant AFM coupling. Accompanying these magnetic changes are a drastic reduction in the electrical resistivity by up to four orders of magnitude. Moreover, the I-V characteristic features a sharp quantum switching that closely tracks the lattice change in the *a* axis, offering a rare negative differential resistance. At the heart of these phenomena is the newly uncovered *a*-axis magnetoelastic effect in bulk $Cr_2Ge_2Te_6$, which presents tantalizing prospects for unique functional materials and devices.

Electric-current-control of structural and physical properties in correlated and spin-orbit coupled materials is an emergent research topic [18]. Our recent work [19-21] indicates that electric current can effectively control quantum states in materials with a strong magnetoelastic coupling and a distorted lattice [19-21]. Note that the small DC current-controlled phenomena are fundamentally different from the Poole-Frenkel effect, which is an increase of the electrical conductivity, consequence of a strong electric field of over $10^5$ V/cm that increases the number of electrons but not their mobility [22]. In addition, studies of current-controlled phenomena require robust, innovative techniques that eliminate effects of Joule heating. In this work, we use a small DC electric current as an external stimulus to control the structural and physical properties in $Cr_2Ge_2Te_6$.

Single crystals of $Cr_2Ge_2Te_6$ were grown using the self-flux technique [11]. The magnetization and electrical resistivity were simultaneously measured using a Quantum Design MPMS-XL magnetometer with a homemade probe [20]. The lattice parameters and I-V characteristics were also culled simultaneously using a Bruker Quest ECO single-crystal diffractometer with a home-developed capability that allows simultaneous measurements of the crystal structure and transport properties as functions of electric current and temperature (**Fig.1a**) [19,20]. The relative permittivity was collected using a QuadTech 7600 LCR with a frequency range of 10Hz - 2MHz and a Quantum Design PPMS-Dynacool with a 14T magnet.

$Cr_2Ge_2Te_6$ crystallizes in the rhombohedral lattice with space group R-3 (No. 148) [13], like its sister compound $Cr_2Si_2Te_6$ [16]. Edge-sharing $CrTe_6$ octahedra form 2D honeycomb planes. Between the planes is an empty space or a van der Waals gap (~ 3.35 Å, slightly smaller than 3.40 Å in $Cr_2Si_2Te_6$ [13]) (**Fig.1a**). A key structural element is the planar Cr-Cr bond distance *d* of two neighboring edge-sharing $CrTe_6$ octahedra, which are distorted, resulting in a discernable displacement of the Cr ions in an alternating manner, as marked by the white arrows in **Fig.1a**. The displacement is approximately along the *c* axis, facilitating the magnetic easy axis to be aligned with the *c* axis.

The lattice parameters *a* and *c* axis increase with increasing temperature above 140 K. As the temperature is lowered below 140 K, this trend reverses for the *a* axis but continues for the *c* axis (data in blue in **Figs.1b-1c**). Below 140 K, the *a* axis and thus the planar Cr-Cr bond distance, *d*, increases with decreasing temperature, exhibiting approximately a linear negative thermal expansion. Consequently, the increased *d* (**Fig.1a**) weakens the direct AFM exchange interaction, facilitating the eventual occurrence of the FM state at $T_C$ (= 66 K). This behavior signals a magnetoelastic coupling along the *a* axis.

The *a* and *c* axis are both susceptible to an applied electric current, but they behave in an opposite manner. Upon the application of current, the *a* axis decreases initially and then gradually levels off with increasing current density J (data in red in **Fig.1b**), whereas the *c* axis increases with increasing J (data in red in **Fig.1c**). As a result, the basal plane shrinks but the *c* axis expands (schematics in **Figs.1b-1c**). In particular, the planar Cr-Cr bond distance *d* is shortened by ~ 0.23%. It needs to be pointed out that the current dependence of the *a* axis sharply contrasts with the temperature dependence of the *a* axis (data in blue in **Figs.1b**), ruling out any significant role of Joule heating [21]. Also, application of current along either the *a* or *c* axis generates similar effects, thus all data presented here correspond to the current applied along the *a* axis.

Before discussing the current-driven physical phenomena, let us first turn to the relative permittivity, $\varepsilon_r$, of $Cr_2Ge_2Te_6$. As shown in **Fig.2**, $\varepsilon_r$ is unusually large (up to 2500) at high temperatures (above room temperature) and decreases with decreasing temperature, eventually leveling off below 140 K, and surprisingly with no sign of the Curie-Weiss dependence, which is normally expected in dielectrics. This peculiar behavior invokes a study of the sister compound $Cr_2Si_2Te_6$ with $T_C$ = 33 K [6], in which infrared spectroscopy and thermal conductivity along with complementary lattice dynamics calculations indicate that the high temperature paramagnetic phase is characterized by strong spin-lattice interactions that give rise to both glassy-behavior and a strong magnetoelastic coupling above $T_C$ [6]. Here the large value of $\varepsilon_r$ along with the unusual temperature dependence above 140 K manifests an extraordinarily soft lattice mode ($^2E_\mu$, infrared active) consistent with the existence of a glassy state. Note that both $Cr_2Ge_2Te_6$ and $Cr_2Si_2Te_6$ share the same crystal and magnetic structures [13]. Moreover, application of a magnetic field along the *a* axis or the magnetic hard axis causes a giant magnetoelectric effect, defined as [$\varepsilon_r$(H) - $\varepsilon_r$(0T)]/$\varepsilon_r$(0T) or $\Delta\varepsilon_r/\varepsilon_r$(0T). At $\mu_oH$ = 14 T, $\Delta\varepsilon_r/\varepsilon_r$(0T) = - 53% (**Fig.2a**). Intriguingly, this value is essentially zero when H is applied along the *c* axis, the magnetic easy axis (**Fig.2b**). The contrasting behavior of $\Delta\varepsilon_r/\varepsilon_r$(0T) for H|| *a* and *c* axis reinforces the strong *a*-axis magnetoelastic coupling, already evident at the aforementioned temperature and current dependence of the *a* axis (**Fig.1b**). With decreasing temperature, the spins are gradually locked in with an increasingly stiffened lattice via the magnetoelastic coupling, rendering the rapid reduction in $\varepsilon_r$ below 140 K (**Fig.2**) as the *a* axis starts to expand (data in blue in **Fig.1b**), aiding the FM state to eventually emerge. The dissipation factor, DF, remains small except in the range of 90-140 K where loss becomes nonnegligible (right scale in **Fig.2**). Nevertheless, the soft lattice mode and the *a*-axis magnetoelastic coupling unveiled in **Fig.2** provides an important insight into the phenomena discussion below.

Now we discuss the current-controlled magnetic properties. As already established, $Cr_2Ge_2Te_6$ ferromagnetically orders at $T_C$ = 66 K with an anisotropy between the *a*- and *c*-axis magnetization, $M_a$ and $M_c$ (**Fig.3a**). The peak in $M_a$ right below $T_C$ implies a possible canted magnetic structure within the basal plane. $Cr_2Ge_2Te_6$ hosts the $Cr^{3+}(3d^3)$ ions, and the Hund's rule coupling renders a half-occupation of each of the three $t_{2g}$ orbitals, giving rise to a maximized spin state, S = 3/2, and an anticipated saturation moment, $M_S$, of 3 $\mu_B$/Cr or 6 $\mu_B$/f.u., which is realized at $\mu_oH$ = 0.16 and 0.43 T for $M_c$ and $M_a$, respectively (Inset in **Fig.3a**). The insulating state is also a result of the half filled $t_{2g}$ orbitals.

Application of electric current readily suppresses $T_C$ and the magnitude of $M_a$ and $M_c$, as illustrated in **Figs. 3b-3c**. A Curie-Weiss analysis indicates a sign change in the Curie-Weiss temperature, $\theta_{CW}$, from + 78 K at J = 0 to – 34 K at J = 2.60 A/cm$^2$ (**Fig.3d**), signaling a change of the ground state. A close examination of $M_a$ and $M_c$ reveals a critical regime of current density of 0.31 < $J_C$ < 0.55 A/cm$^2$, above which a paramagnetic state emerges (shaded area in **Figs. 4a-4b**). The range of $J_C$ is likely due to the magnetic anisotropy between $M_a$ and $M_c$. It is revealing that these magnetic changes closely track the current-induced lattice changes in the planar Cr-Cr bond distance $d$ and the $c$ axis (**Fig.4c**). The rapidly vanishing FM state with J (**Figs.4a-4b**) is clearly a result of the shortened $d$ that enhances the AFM interaction and the thermal fluctuations, and, to a less extent, the elongated $c$ axis that simultaneously weakens the FM superexchange interaction. The same notion also explains the lower $T_C$ in $Cr_2Si_2Te_6$ (32 K) than in $Cr_2Ge_2Te_6$ (66 K). This is because $Cr_2Si_2Te_6$ has a shorter Cr-Cr bond distance $d$ (3.909 Å) than that (3.934 Å) in $Cr_2Ge_2Te_6$ [13], hence a relatively stronger AFM interaction that makes it less favorable for the FM state to stabilize at higher temperatures in $Cr_2Si_2Te_6$.

The $a$-axis electrical resistivity $\rho_a$, which was simultaneously measured with the magnetization, decreases considerably with increasing J, resulting in a reduction by up to four orders of magnitude at J = 5.2 A/cm$^2$ (**Fig.5a**). Moreover, the I-V characteristic shows an "S"-shaped negative differential resistance (NDR) featuring a sharp quantum switching effect (**Fig.5b**).

$Cr_2Ge_2Te_6$ undergoes a sharp switching to a much more metallic state from the insulating state at a threshold voltage $V_{th}$. Interestingly, this I-V curve (blue) closely traces the change in the *a* axis (red) at $V_{th}$ (**Fig.5b inset**) (Note that the I-V curve and the lattice parameters were simultaneously measured). Such a lattice-driven switching effect is rare in bulk materials. The common NDR is a result of either an "electrothermal" effect or a "transferred carrier" effect [23-26], and manifests in "N"-shaped I-V characteristics. The S-shaped NDR is far less common but highly desirable for nonvolatile memory devices [27-29]. Previous studies indicate that so far the S-shaped NDR occurs only in a few bulk materials such as $VO_2$, $CuIr_2S_{4-x}Se_x$ and $1T-TaS_2$ and is attributed to the inherent first-order structural or metal-insulator transition [e.g., 30,31]. Here the sharp switching effect that closely tracks the lattice change occurs in a material with no such first-order transition, suggesting a different mechanism. A similar behavior is also seen in spin-orbit-coupled oxides such as $Sr_2IrO_4$ where the strong spin-orbit interaction renders a strong lattice coupling so that the I-V characteristics closely follows the current-controlled lattice change [18].

    This work demonstrates that application of small DC electric current can effectively control the structural and physical properties in the chalcogenide, and even induce phenomena seldom seen in other materials. At the heart of the tunability is the newly found *a*-axis magnetoelastic coupling in $Cr_2Ge_2Te_6$, which has a distinct soft lattice mode at high temperatures. Our on-going study shows strong evidence that this tunability may be widespread in other related tellurides. Clearly, the current-controlled phenomena in this class of materials pose tantalizing prospects for unique functional materials and devices, but a better understanding of them needs to be first established. In particular, the coupling between electric current and the lattice, orbitals or magnetic moments in these materials has yet to be adequately described.

**Acknowledgement** This work is supported by NSF via grant DMR 1903888. SA acknowledges support of Deutsche Forschungsgemeinschaft (DFG) through Grant AS 523/4-1; BB through SFB 1143 (project-id 247310070).

**Captions**

**Fig.1. Crystal Structure**: **(a)** The crystal structure of $Cr_2Ge_2Te_6$; a schematic highlighting that the Cr ions are noticeably displaced in an alternating manner, as marked by the white arrows in neighboring edge-sharing $CrTe_6$ octahedra; the Cr-Cr bond distance is denoted by *d*, which enables the direct AFM exchange interaction; a photo illustrating the home-developed setup for the simultaneous measurements of the lattice parameters and transport properties as functions of electric current and temperature. **(b)** and **(c)** The electric current (red) and temperature (blue) dependence of the *a* axis and *c* axis lattice parameters. The *b* axis follows the same temperature and current dependence as the *a* axis. The schematics illustrate the current-induced effect on the basal plane and the *c* axis.

**Fig.2. Dielectric Constant**: The temperature dependence of the relative permittivity $\varepsilon_r$ and the dissipation factor DF (thin lines, right scale) with the electric field E applied along the *c* axis and a frequency *f* = 1 MHz when **(a)** the magnetic field H is applied along the *a* axis or the magnetic hard axis and **(b)** H is applied along the *c* axis, the magnetic easy axis. Note that the magnetodielectric effect is -53% when H∥*a* and zero when H∥*c*.

**Fig. 3. Magnetic Properties**: The temperature dependence of (a) the a- and c-axis magnetization $M_a$ and $M_c$ at 0.1 T, **(b)** $M_a$ with current density J∥*a*, **(c)** $M_c$ with J∥*a*, and **(d)** the reciprocal magnetic susceptibility $\Delta\chi^{-1}$ ($\Delta\chi = \chi - \chi_o$, where $\chi_o$ is temperature-independent magnetic susceptibility). Inset in (a): the isothermal M(H) at 1.8 K.

**Fig.4. Correlation between Magnetic and Structural Properties**: The current density J dependence of **(a)** the Curie temperature $T_C$ retrieved from $M_a$ and $M_c$, **(b)** the magnitude of $M_a$ and $M_c$ at 10 K, and **(c)** the planar Cr-Cr bond distance (red) and the c axis (blue) at 80 K. Inset: highlighting *d* in the $CrTe_6$ octahedra.

**Fig.5. Transport Properties**: **(a)** The temperature dependence of the *a*-axis resistivity $\rho_a$ at various current density J. Note that $\rho_a$ was simultaneously measured with $M_a$. **(b)** The I-V characteristic (blue) and the simultaneously measured *a* axis (red) at 80 K. Inset: Highlighting the direct correlation between the I-V curve and the *a* axis near the switching threshold $V_{th}$.

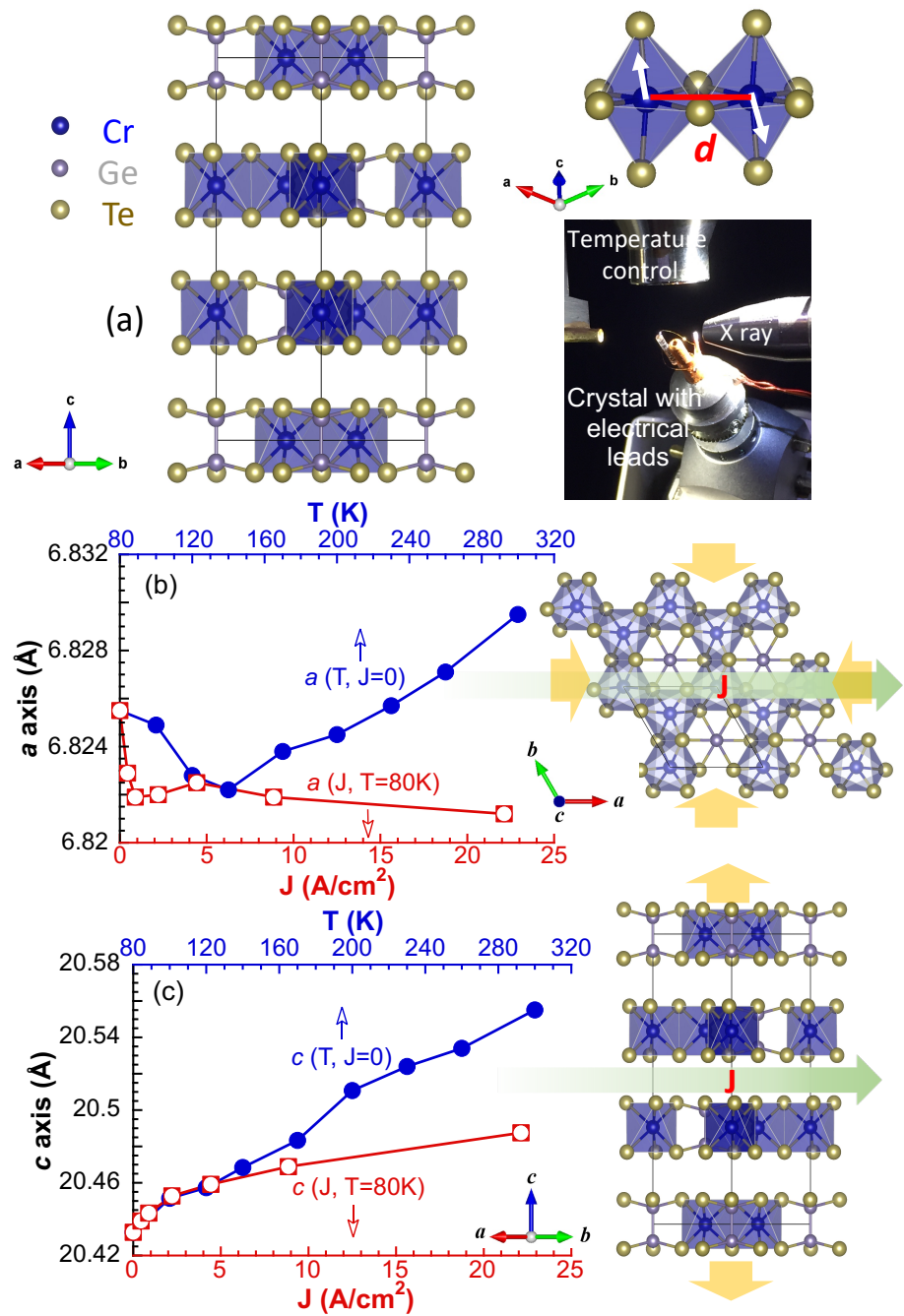

Figure 1

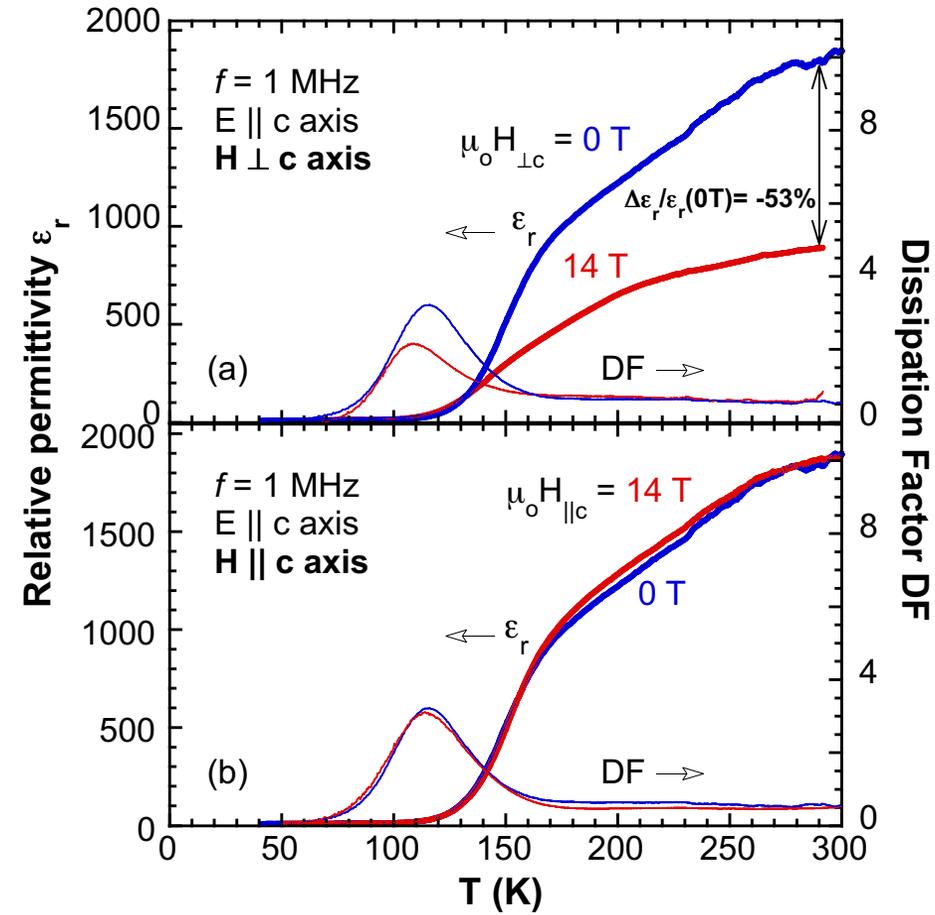

Figure 2

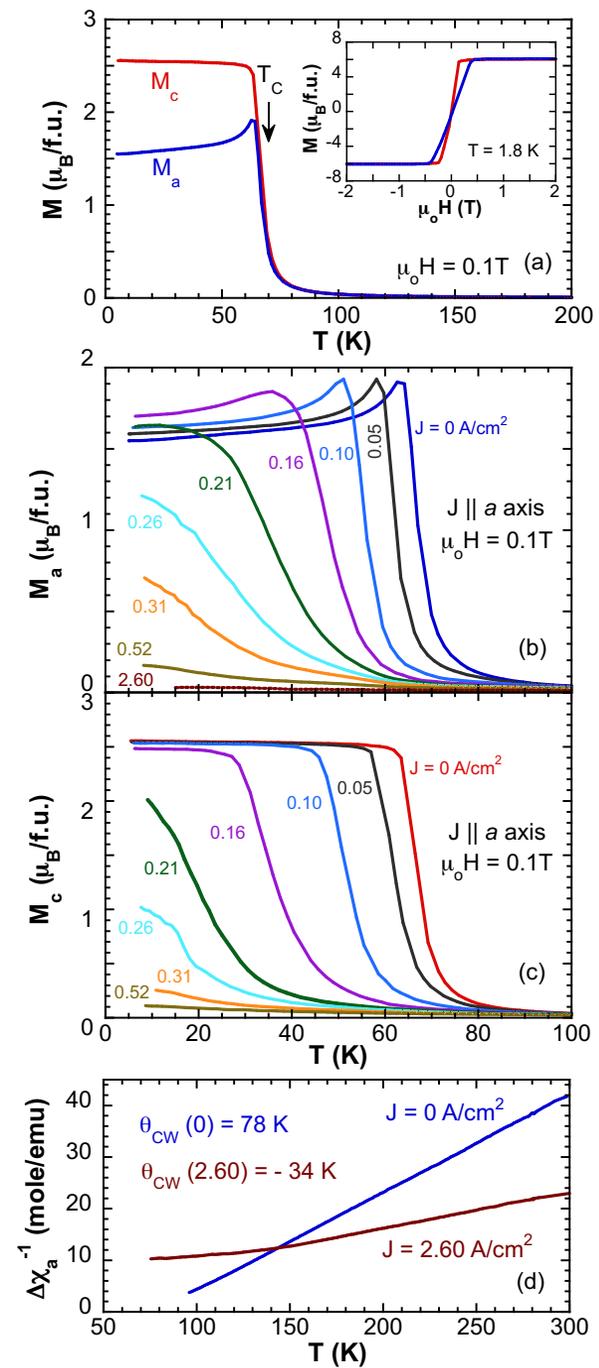

Figure 3

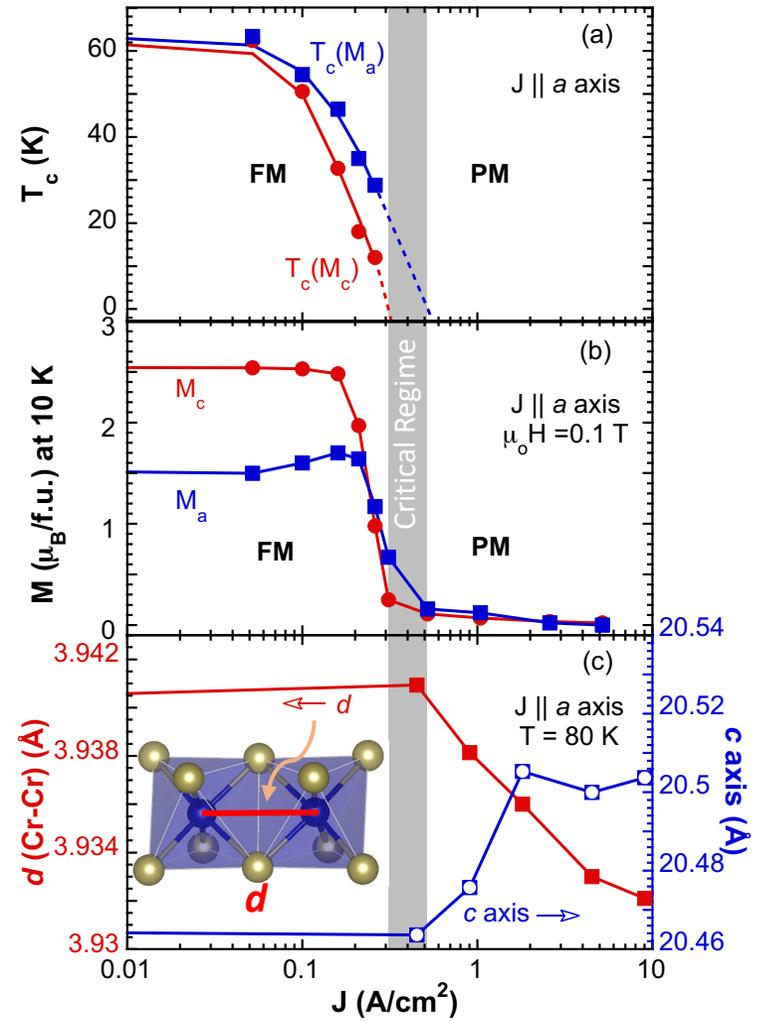

Figure 4

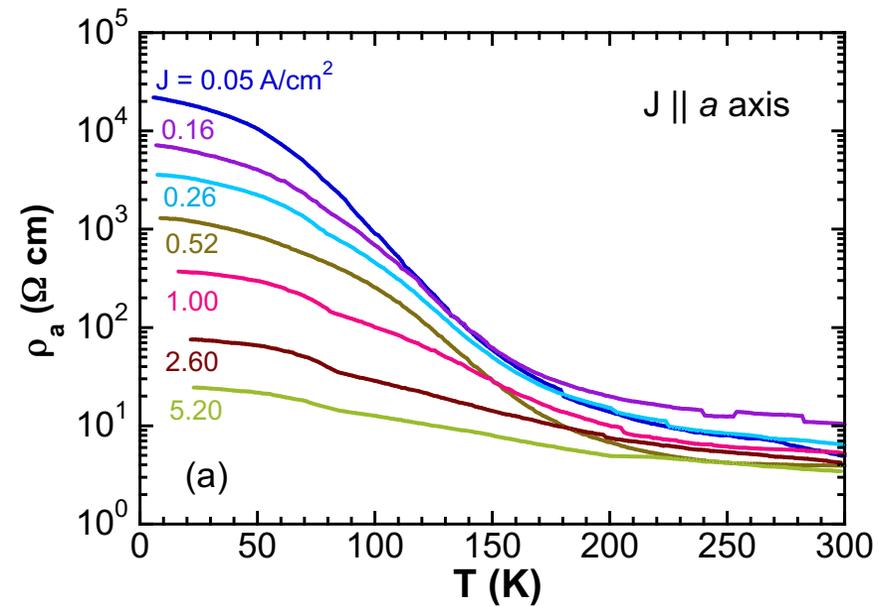
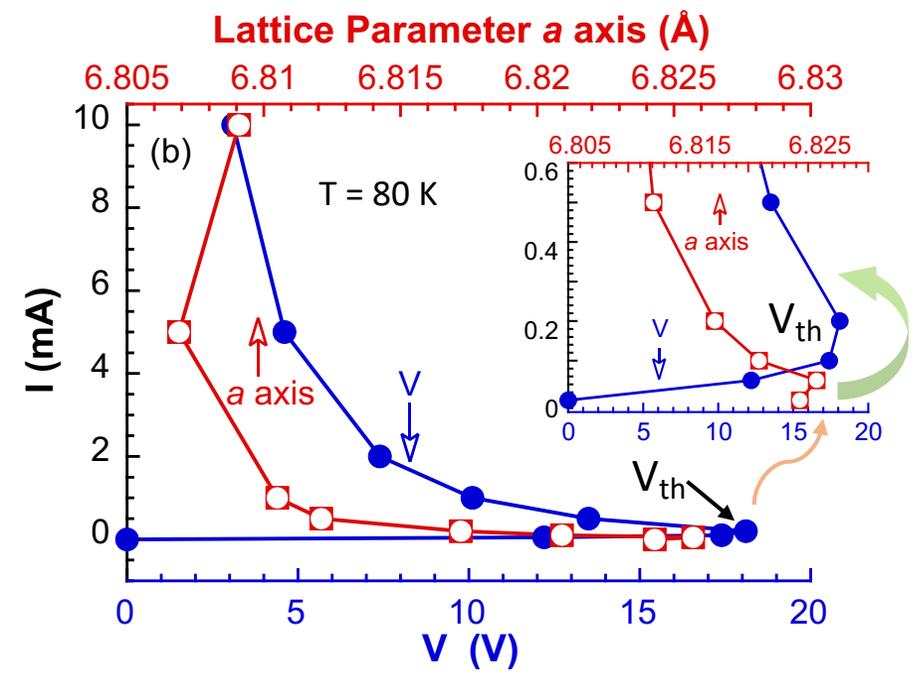

Figure 5